\documentstyle[aps,epsf,amssymb]{revtex}

\begin{document}

\newcommand{\ini}{\begin{equation}}
\newcommand{\fin}{\end{equation}}
\newcommand{\inir}{\begin{eqnarray}}
\newcommand{\finr}{\end{eqnarray}}
\newcommand{\inif}{\begin{figure}}
\newcommand{\finf}{\end{figure}}
\newcommand{\bc}{\begin{center}}
\newcommand{\ec}{\end{center}}
\def\ol{\overline}
\def\pa{\partial}
\def\ra{\rightarrow}
\def\ts{\times}
\def\df{\dotfill}
\def\bs{\backslash}
\def\dg{\dagger}

$~$

\hfill DSF-20/2000

\vspace{1 cm}

\centerline{{\bf TALK ON FERMION MASSES AND MIXINGS
\footnote{Presented at the Instituto Superior Tecnico, Lisbon, May 2000}}}

\vspace{1 cm}

\centerline{\large{D. Falcone}}

\vspace{1 cm}

\centerline{Dipartimento di Scienze Fisiche, Universit\`a di Napoli,}
\centerline{Complesso di Monte Sant'Angelo, Via Cintia, Napoli, Italy}

\centerline{{\tt e-mail: falcone@na.infn.it}}

\vspace{1 cm}

\begin{abstract}

\noindent
Two topics are covered in this paper. In the first part the relation
between quark mass matrices and observable quantities in gauge theories.
In the second part neutrino masses and mixings in a seesaw framework.

\end{abstract}

\newpage

\twocolumn

\noindent
This talk is divided in two parts: Part I deals with quark mass matrices
and observable quantities and is based on three papers \cite{f1,f2,f3}.
Part II deals with neutrino masses and mixings in
a seesaw framework and is based on another three papers \cite{f4,f5,f6}. 

$~$

\noindent
{\bf Part I - Quarks}

$~$

In the standard model (SM), which is a gauge theory with
$SU(3)_c \ts SU(2)_L \ts U(1)_Y$ as a symmetry group, let us consider the
quark mass and charged weak current terms in the Lagrangian:
\ini
\ol{u}_L M_u u_R+\ol{d}_L M_d d_R+ g \ol{u}_L d_L W.
\fin
When we diagonalize both mass matrices by biunitary transformations, we get
the terms
\ini
\ol{u}_L D_u u_R+\ol{d}_L D_d d_R+ g \ol{u}_L V_{CKM} d_L W,
\fin
where $V_{CKM}$ is the quark mixing matrix \cite{ckm}.
There is a clear hierarchy of quark masses: $m_u \ll m_c \ll m_t$ and
$m_d \ll m_s \ll m_b$ ($m_b \ll m_t$). Moreover, $V_{CKM}$ is near the
identity and $V_{ub} \ll V_{cb} \ll V_{us}$.

The following (unitary)
transformations have no physical consequences, that is they do not
change masses and mixings:
\ini
u_L \ra U u_L,~d_L \ra U d_L
\fin
\ini
u_R \ra V_u u_R,~d_R \ra V_d d_R.
\fin
As you see, the left-handed states transform in the same way, while the
right-handed states may undergo different transformations.
The two mass matrices
$M_u$, $M_d$ have 36 real parameters while $D_u$, $D_d$, $V_{CKM}$
have 10 real parameters (observable quantities). By using transformations
(3),(4) we can go to a basis where the mass matrices contain exactly 10
real parameters. This is called a minimal parameter basis (MPB)\cite{koide}. 

However, by using transformations (3),(4) we can get also
$M_u$, $M_d$ both hermitian \cite{fj}, or $M_u$, $M_d$ both in the
nearest neighbor interaction (NNI) form
\cite{blm}, or $M_u$ diagonal and $M_d$ hermitian or containing three zeros
\cite{ma}. 
In fact, we can diagonalize $M_u$ by a biunitary transformation, and use the
freedom in $V_d$ to get $M_d$ hermitian or containing three zeros. There
are 54 bases with three zeros, out of 84 possibilities. In this way
$M_u$ has three real parameters and $M_d$ seven real parameters, six moduli
and one phase (but keeping three phases preserves an arbitrary representation
for $V_{CKM}$). 

As an example of a MPB with $M_u$ diagonal and $M_d$ with three zeros,
we can consider the one studied in
ref.\cite{f1}, where
\ini
M_d \simeq \left( \begin{array}{ccc}
0 & \sqrt{m_d m_s} & 0 \\ \sqrt{m_d m_s} & m_s e^{i \varphi} & m_s \\
0 & m_b/\sqrt5 & 2 m_b/\sqrt5
\end{array} \right).
\fin
There are three simple relations among its elements, and the mixings can be
written in terms of down quark masses:
\inir
V_{us} & \simeq & \sqrt{\frac{m_d}{m_s}} \\
V_{cb} & \simeq & \frac{3}{\sqrt5} {\frac{m_s}{m_b}} \\
V_{ub} & \simeq & \frac{1}{\sqrt5} \sqrt{\frac{m_d m_s}{m_b}}.
\finr
Therefore on this basis we have seven independent real parameters (six masses
and one phase), instead of ten. We can check also that
$V_{us} V_{cb} \simeq 3 V_{ub}$. The relation $|M_{d22}| \simeq M_{d23}$
is due to the value $\delta \simeq 75^{\circ}$ in $V_{CKM}$.

Other interesting bases, with $M_u$ diagonal and $M_d$ triangular, are in
refs.\cite{hs,kmw}. In particular, from ref.\cite{kmw} we get the form
\ini
|M_d| \simeq \left( \begin{array}{ccc}
m_d & m_s V_{us} & m_b V_{ub} \\ 0 & m_s & m_b V_{cb} \\ 0 & 0 & m_b
\end{array} \right).
\fin

Non diagonal bases include hermitian matrices with no zeros on
the diagonal \cite{bbhl}. The physical content of such bases can be explained
considering that, for example, solution 3 by Ramond, Roberts, Ross \cite{rrr}
is obtained setting $M_{u11}=M_{d11}=0$ in one of them. Moreover, we have also
the basis with hermitian matrices and $M_{u13}=M_{d13}=0$ \cite{fx}, or
$M_{u11}=M_{d11}=M_{d13}=0$ \cite{beg}, and the physical content is given
by the ansatz with $M_{11}=M_{13}=0$ in both matrices (this is contraddicted
in ref.\cite{ckw}, where other hermitian mass matrices are studied and
a non parallel structure is found).

Let us now turn to the left-right model (LRM), based on the symmetry
$SU(3)_c \ts SU(2)_L \ts SU(2)_R \ts U(1)$, and consider the part of the
Lagrangian containing the quark mass terms and the charged currents:
\ini
\ol{u}_L M_u u_R+\ol{d}_L M_d d_R+ g_L \ol{u}_L d_L W_L+g_R \ol{u}_R d_R W_R.
\fin
When we diagonalize the mass matrices we obtain
\ini
\ol{u}_L D_u u_R+\ol{d}_L D_d d_R+ g_L \ol{u}_L V_L d_L W_L+
g_R \ol{u}_R V_R d_R W_R,
\fin
with two mixing matrices $V_L$, $V_R$, and $V_L=V_{CKM}$. The transformations
that do not change masses and mixings are now
\ini
u_L \ra U u_L,~d_L \ra U d_L
\fin
\ini
u_R \ra V u_R,~d_R \ra V d_R.
\fin
Thus, also the right-handed fields transform in the same way, and if we diagonalize
$M_u$ by a biunitary transformation, then $M_d$ is fixed: The 54 bases
of the SM become strong ansatze in the LRM. To select them we have to look
at right-handed mixings \cite{rh}.

As an example, consider the 2-generation case, where we have four SM bases,
according to where we put the zero in $M_d$ (taking $M_u=D_u$, $\lambda=0.22$):
\ini
M_d \simeq \left( \begin{array}{cc}
           0 & \sqrt{m_d m_s} \\ \sqrt{m_d m_s} & m_s
\end{array} \right),
V_R \simeq \left( \begin{array}{cc}
           -1 & \lambda \\ \lambda & 1
\end{array} \right);
\fin
\ini
M_d \simeq \left( \begin{array}{cc}
           \sqrt{m_d m_s} & 0 \\ m_s & \sqrt{m_d m_s} 
\end{array} \right),
V_R \simeq \left( \begin{array}{cc}
           \lambda & 1 \\ -1 & \lambda
\end{array} \right);
\fin
\ini
M_d \simeq \left( \begin{array}{cc}
           -m_d & \sqrt{m_d m_s} \\ 0 &  m_s
\end{array} \right),
V_R \simeq \left( \begin{array}{cc}
           -1 & 0 \\ 0 & 1
\end{array} \right);
\fin
\ini
M_d \simeq \left( \begin{array}{cc}
           \sqrt{m_d m_s} & -m_d \\ m_s & 0
\end{array} \right),
V_R \simeq \left( \begin{array}{cc}
           0 & 1 \\ -1 & 0
\end{array} \right).
\fin

In the 3-generation case we rely on the following considerations.
A recent analysis, by T. Rizzo \cite{riz}, of right-handed currents in B
decay, within the LRM, suggests that $V_{cb}^R$ is large and perhaps near unity.
From inclusive semileptonic decays of B mesons one has
$V_{cb}^R \gtrsim 0.782$. Moreover, if as suggested by Voloshin \cite{vol},
right-handed currents can help to solve the B semileptonic branching fraction
and charm counting problems, then $V_{cb}^R \ge 0.908$. We begin our
selection by using $V_{cb}^R \ge 0.750$. Setting $M_u=D_u$, we denote elements
in $M_d$ by 
$$
\left( \begin{array}{ccc}
1 & 2 & 3 \\ 4 & 5 & 6 \\ 7 & 8 & 9
\end{array} \right).
$$
There are 16 ansatze out of 54 that satisfy our bound. Ten of them
are in the following table. We have excluded those giving the particular strange result
that some element in $M_d^{\dg} M_d = V_R D_d^2 V_R^{\dg}$ is exactly zero.

\begin{center}
\begin{tabular}{|c|cccccccccc|}\hline
zeros & 124 & 236 & 146 & 256 & 127 & 128 & 479 & 589 & 467 & 568 \\
$V_{cb}^R$ & .896 & .896 & .789 & .789 & .914 & .914 & .999 & .999
& .871 & .871 \\ \hline
\end{tabular}
\end{center}

We give now an example of a successful ansatz, namely 124, with
\ini
|M_d|=\left( \begin{array}{ccc}
      0 & 0 & 0.023 \\ 0 & 0.106 & 0.104 \\ 0.541 & 2.687 & 1.213
\end{array} \right),
\fin
\ini
|V_R| \simeq \left( \begin{array}{ccc}
          1 & \lambda & \lambda \\
          \lambda & 2 \lambda & 1 \\
         \lambda & 1 & 2 \lambda
\end{array} \right).
\fin

Further constraints on the form of $V_R$ come from the $K_L-K_S$ mass difference
and $B-\ol{B}$ mixing, reported in ref.\cite{riz}. Models 128, 479, 589 are
reliable, for example 128:
\ini
|M_d|=\left( \begin{array}{ccc}
      0 & 0 & 0.023 \\ 0.103 & 0.021 & 0.104 \\ 2.741 & 0 & 1.213
\end{array} \right),
\fin
\ini
|V_R| \simeq \left( \begin{array}{ccc}
          \lambda & 1 & 2 \lambda \\
          2 \lambda^2 & 2 \lambda & 1 \\
         1 & \lambda & \lambda^5
\end{array} \right).
\fin
Of course, other ansatze can be obtained starting from a diagonal $M_d$.
We stress the simple result that, if $V_{cb}^R$ is really large, then
hermitian or symmetric mass matrices are not reliable (remember: for
hermitian or symmetric matrices $|V_R|=|V_L|$). Notice also that
non symmetric mass matrices have important applications in the leptonic sector
in connection with the large mixing of neutrinos \cite{non}.

By using transformations (12),(13) it is possible to change the structure
of both $M_u$ and $M_d$. Although such forms can be more interesting
to discover an underlying theory of fermion masses and mixings, they lead
to the same observable parameters in LRM, and we need other observable
quantities to make a selection of such models with non diagonal mass matrices.
These new physical parameters exist in extensions of the LRM (for example
the $SO(10)$ model). We have here simply attempted to begin with a systematic
study, within the LRM, of quark mass matrices which have a general form in
the SM. For effects of right-handed mixings in $SO(10)$ and proton decay
see ref.\cite{achi}.

$~$

\noindent
{\bf Part II - Neutrinos}

$~$

Let us start from the 1-generation seesaw, where the full neutrino mass
matrix is given by
\ini
\left( \begin{array}{cc}
0 & m_D \\ m_D & M_R
\end{array} \right),
\fin
with $m_D$ the Dirac mass and $M_R$ the right-handed Majorana mass.
Assuming $m_D \ll M_R$, we get a small eigenvalue $m_L \simeq m_D^2/M_R$ and
a big one, $m_R \simeq M_R$. Usually $M_D \sim m_q$ or $M_D \sim m_l$, with
$m_q$ a quark mass and $m_l$ a charged lepton mass.
In the 3-generation seesaw, we have the matrix
\ini
\left( \begin{array}{cc}
0 & M_D \\ M_D & M_R
\end{array} \right),
\fin
with $M_D$ and $M_R$ $3 \ts 3$ matrices (we assume $M_D$ to be real symmetric),
and an effective light neutrino mass matrix
\ini
M_L \simeq M_D M_R^{-1} M_D
\fin
which can be inverted to give
\ini
M_R \simeq M_D M_L^{-1} M_D.
\fin

We aim at calculating the scale and the form of $M_R$, because in unified
models, where neutrino mass is most natural ($SO(10)$), this scale is
intermediate (around $10^{11}$ GeV)
in the non supersymmetric case and of unification ($10^{16}$ GeV) in the
supersymmetric case \cite{dkp}.

We rely on quark-lepton symmetry, which in this context means that
\ini
M_D \simeq \frac{m_{\tau}}{m_b}$diag$ (m_u,m_c,m_t),
\fin
and $M_l$, $M_d$, $M_u$ also nearly diagonal (see for example ref.\cite{rrr}).
In such a case we can soon obtain $M_L$ from experimental data, by means
of the formula
\ini
M_L \simeq U D_L U^T
\fin
with $D_L =~$diag$ (m_1,m_2,m_3)$ and the lepton mixing matrix $U$
\cite{mns} given by
\cite{akh}
\ini
U=\left( \begin{array}{ccc}
c & s & \epsilon \\
-\frac{1}{\sqrt2}(s+c\epsilon) & \frac{1}{\sqrt2}(s-c\epsilon) &
\frac{1}{\sqrt2} \\
\frac{1}{\sqrt2}(s-c\epsilon) & -\frac{1}{\sqrt2}(s+c\epsilon) &
\frac{1}{\sqrt2}
\end{array} \right)
\fin
where $\epsilon$ is small ($\epsilon^2 \lesssim 0.03$).
This form includes the results by
Chooz and SuperKamiokande. Since we do not know neutrino masses and mixing
so well, our approximations can be considered good.

From neutrino oscillation data we have for solar neutrinos
\cite{bks}
\ini
\Delta m^2_{sol} \sim 10^{-6} eV^2 (SMA~MSW)
\fin
\ini
\Delta m^2_{sol} \sim 10^{-5} eV^2 (LMA~MSW)
\fin
\ini
\Delta m^2_{sol} \sim 10^{-10} eV^2 (VO)
\fin
and for atmospheric neutrinos
\ini
\Delta m^2_{atm} \sim 10^{-3} eV^2.
\fin
Then it is clear that $\Delta m^2_{atm} \gg \Delta m^2_{sol}$. We can set
\cite{af}
\ini
\Delta m^2_{sol}=m_2^2-m_1^2,~\Delta m^2_{atm}=m_3^2-m_{1,2}^2
\fin
and assuming $m_3 >0$ there are three possible spectra
for light neutrinos \cite{af}:
\ini 
m_3 \gg |m_2|,|m_1|~~(hierarchical)
\fin
\ini
|m_1| \sim |m_2| \gg m_3 ~~(inverted~hierarchy)
\fin
\ini
|m_1| \sim |m_2| \sim m_3 ~~(nearly~degenerate).
\fin
We work in a CP conserving framework, but we allow for negative Majorana
masses. In a more general approach, two phases in $U$ or $D_L$
connect among the relative signs (see for example ref.\cite{gg}).
For the sine $s$ of the solar mixing angle one has $s \simeq 0$ in the
SMA MSW and $s \simeq 1/\sqrt2$ in the LMA MSW or VO. 

$Hierarchical~spectrum$. If there was no mixing at all ($U=I$),
the scale of $M_R$ would be
$$
M_{R33} \sim \frac {k^2 m_t^2}{m_3},
$$
with $k=m_{\tau}/m_b$. For $s \simeq 0$, corresponding to the SMA solution,
we get
\ini
M_L^{-1} \simeq \left( \begin{array}{ccc}
1/m_1 & 0 & 0 \\ 0 & 1/2m_2 & -1/2m_2 \\ 0 & -1/2m_2 & 1/2m_2
\end{array} \right),
\fin
when $\epsilon^2 m_3 \ll m_1$, and the scale is given by
\ini
M_{R33} \sim \frac{1}{2} \frac {k^2 m_t^2}{m_2},
\fin
which is greater or equal to $10^{15}$ GeV, that is near the unification
scale. The leading form for $M_R$ is
\ini
M_R \sim ~$diag$ (0,0,1),
\fin
similar to the leading form for $M_D$. However, if $\epsilon^2 m_3 \simeq m_1$,
then $M_{R33}$ goes to zero and we have a different structure for $M_R$.
For $s \simeq 1/\sqrt2$ we consider three subcases:
$|m_2| \gg |m_1|$, $m_2 \simeq m_1$ and $m_2 \simeq -m_1$.

{\bf 1.} If there is full hierarchy, we have
\ini
M_L^{-1} \simeq \left( \begin{array}{ccc}
1 & -1/\sqrt2 & 1/\sqrt2 \\ -1/\sqrt2 & 1/2 & -1/2 \\ 1/\sqrt2 & -1/2 & 1/2
\end{array} \right) \frac{1}{2m_1}
\fin
and the scale is
\ini
M_{R33} \sim \frac{1}{4} \frac {k^2 m_t^2}{m_1}
\fin
which gives $10^{16}$ GeV or more in the LMA case and $10^{18}$ GeV or more
in the VO case (towards the Planck scale). Again the leading form for $M_R$
is hierarchical and diagonal, reflecting the hierarchy of Dirac masses.

{\bf 2.} If $m_2 \simeq m_1$, then
\ini
M_L^{-1} \simeq \left( \begin{array}{ccc}
2 & 0 & 0 \\ 0 & 1 & -1 \\ 0 & -1 & 1
\end{array} \right) \frac{1}{2m_2},
\fin
assuming $\epsilon^2 m_3 \ll m_2$, and the scale is
\ini
M_{R33} \sim \frac{1}{2} \frac {k^2 m_t^2}{m_2},
\fin
which gives $10^{15}$ GeV or more, near or above the unification scale.
The leading $M_R$
is hierarchical and diagonal, unless $\epsilon^2 m_3 \simeq m_2$, when
$M_{R33}$ goes to zero.

{\bf 3.} For $m_2 \simeq -m_1$ there are some possibilities:
$$
M_L^{-1} \simeq \left( \begin{array}{ccc}
0 & -\sqrt2 / m_2 & \sqrt2 / m_2 \\
-\sqrt2 / m_2 & -1/2m_3 & -1/2m_3 \\
\sqrt2 / m_2 & -1/2m_3 & -1/2m_3
\end{array} \right),
$$
$$
M_L^{-1} \simeq \left( \begin{array}{ccc}
0 & -1/\sqrt2 & 1/\sqrt2 \\
-1/\sqrt2 & 0 & 0 \\
1/\sqrt2 & 0 & 2 \epsilon
\end{array} \right) \frac{1}{m_2},
$$
giving $M_R$ around the unification scale. An interesting case is $\epsilon m_3 \simeq -m_2$,
when $m_2 <0$, $m_1,m_3 >0$, and
\ini
M_L^{-1} \simeq \left( \begin{array}{ccc}
0 & -1/\sqrt2 & 1/\sqrt2 \\
-1/\sqrt2 & 2 \epsilon & 0 \\
1/\sqrt2 & 0 & 0
\end{array} \right) \frac{1}{m_2},
\fin
with the leading form
\ini
M_R \sim \left( \begin{array}{ccc}
0 & 0 & 1 \\ 0 & 0 & 0 \\ 1 & 0 & 0
\end{array} \right),
\fin
and a scale given by the element
\ini
M_{R13} \sim \frac{1}{\sqrt2} \frac{k^2 m_u m_t}{m_2}
\fin
yielding $10^{11}$ GeV or above, near the intermediate scale of a
$SO(10)$ model.
This last form is similar to the one studied in ref.\cite{bs},
\ini
M_R \sim  \left( \begin{array}{ccc}
0 & \sigma^2 & 1 \\ \sigma^2 & \sigma^2 & 0 \\ 1 & 0 & 0 
\end{array} \right) M_0,
\fin
with $\sigma^2 =m_c/m_t$ and $M_0=10^{12}$ GeV. For other forms at the
intermediate scale
see refs.\cite{js,abr,ab}.

For inverted hierarchy, generally we have
\ini
M_{R33} \sim \frac{k^2 m_t^2}{m_3}
\fin
and the hierarchical and diagonal leading form with a scale at the
unification scale or above.

For nearly degenerate spectrum with $m_0 \simeq 2$ eV, the scale is intermediate
and depends on $1/m_0$. However, in this case there are instabilities
with respect to small entries \cite{akh}.
Moreover, this spectrum is hard to obtain in the seesaw mechanism
\cite{af}.

We give now a summary on neutrinos.
There are few leading forms for $M_R$. The diagonal form
\ini
M_R \sim \left( \begin{array}{ccc}
0 & 0 & 0 \\ 0 & 0 & 0 \\ 0 & 0 & 1 
\end{array} \right)
\fin
leads to the unification scale or above; in particular, for the VO solution
and full hierarchy we go near the Planck scale. The off-diagonal forms
\cite{f6,ab}
\ini
M_R \sim \left( \begin{array}{ccc}
0 & 0 & 1 \\ 0 & 0 & 0 \\ 1 & 0 & 0 
\end{array} \right),
~\left( \begin{array}{ccc}
0 & 0 & 0 \\ 0 & 0 & 1 \\ 0 & 1 & 0 
\end{array} \right),
\fin
are consistent with the intermediate scale of non supersymmetric models.
In such cases the structure of $M_R$ is very different from that of
$M_D \sim ~$diag$ (0,0,1)$.
From the point of view of the effective parameters $m_i,\epsilon$ this is
due to some suitable cancellations, but there could be an underlying theory.
Of course, the relation $M_D \sim M_l$ reduces scales by almost three orders
\cite{f4},
so the unification scale can become the intermediate scale.

We finish with a brief comment on neutrinos and dark matter.
From inflation and CMB experiments \cite{cmb} we can argue that $\Omega=1$.
The old paradigm about the content of the universe was
$\Omega_B \simeq 0.1$ (baryons), $\Omega_{HDM}=\Omega_{\nu} \simeq 0.2$
(hot dark matter),
$\Omega_{CDM} \simeq 0.7$ (cold dark matter)
and $\Omega_{\Lambda} = 0$ (vacuum energy). New results from high-z
supernovae and other hints lead to $\Omega_{\Lambda} \simeq 0.7$
\cite{tur}.
Then
$\Omega_{CDM} \simeq 0.2$ for structure formation and we are left with
$\Omega_{\nu} \simeq 0$. Thus, we would like to stress that the hierarchical spectrum
for light neutrinos leads just to $\Omega_{\nu} \simeq 0$, while
the degenerate spectrum to the old position $\Omega_{\nu} \simeq 0.2$
\cite{gg}.
Therefore, it seems that the new results support the hierarchical spectrum.

$~$

The author thanks the Instituto Superior Tecnico for hospitality.

\end{document}